\documentclass[compress,sort,final,11pt,3p]{elsarticle}

\usepackage[colorlinks]{hyperref}
\usepackage{amsmath,amssymb}
\usepackage{upgreek}
\usepackage{multirow}
\usepackage{tablefootnote}
\usepackage{listings}
\usepackage{xcolor}
\usepackage{booktabs}
\usepackage{xspace}
\usepackage{cleveref}
\usepackage{upquote}
\usepackage[utf8]{inputenc}

\journal{Computer Physics Communications}

\usepackage[framemethod=TikZ]{mdframed}
\usepackage{tikz} 

\definecolor{maroon}{cmyk}{0, 0.87, 0.68, 0.32}
\definecolor{halfgray}{gray}{0.55}
\definecolor{slha_frame}{RGB}{207, 207, 207}
\definecolor{slha_bg}{RGB}{247, 247, 247}
\definecolor{slha_red}{RGB}{186, 33, 33}
\definecolor{slha_green}{RGB}{0, 128, 0}
\definecolor{slha_cyan}{RGB}{64, 128, 128}
\definecolor{slha_purple}{RGB}{170, 34, 255}

\definecolor{mathematica_frame}{RGB}{207, 207, 207}
\definecolor{mathematica_bg}{RGB}{247, 247, 247}
\definecolor{mathematica_red}{RGB}{186, 33, 33}
\definecolor{mathematica_green}{RGB}{0, 128, 0}
\definecolor{mathematica_cyan}{RGB}{64, 128, 128}
\definecolor{mathematica_purple}{RGB}{170, 34, 255}

\lstnewenvironment{MIN}[1][]{%
\mdframed[roundcorner=5pt,backgroundcolor=mathematica_bg]
  \renewcommand{\thelstnumber}{In[\arabic{lstnumber}]}
  \lstset{language=MathIn,numbers=left,basicstyle=\ttfamily\footnotesize,#1,escapeinside=||}%
}{%
\endmdframed
}

\lstnewenvironment{MOUT}[1][]{%
\mdframed[roundcorner=5pt,backgroundcolor=mathematica_bg]
  \renewcommand{\thelstnumber}{Out[\arabic{lstnumber}]}
  \lstset{language=MathOut,numbers=left,basicstyle=\ttfamily\footnotesize,#1}%
}{%
\endmdframed
}

\usepackage{listings}
\lstdefinelanguage{Fortran90}{
    morekeywords={Real,Complex,Intent},%
    emph={End,Subroutine,dp,in,Function,Implicit,None},%
    emphstyle={\color{mathematica_purple}},    
    sensitive=true,%
    morecomment=[l]\#,%
    morestring=[b]',%
    morestring=[b]",%
    morestring=[s]{'''}{'''},% used for documentation text (mulitiline strings)
    morestring=[s]{"""}{"""},% added by Philipp Matthias Hahn
    morestring=[s]{r'}{'},% `raw' strings
    morestring=[s]{r"}{"},%
    morestring=[s]{r'''}{'''},%
    morestring=[s]{r"""}{"""},%
    morestring=[s]{u'}{'},% unicode strings
    morestring=[s]{u"}{"},%
    morestring=[s]{u'''}{'''},%
    morestring=[s]{u"""}{"""},%
    identifierstyle=\color{black}\ttfamily,
    commentstyle=\color{slha_cyan}\ttfamily,
    stringstyle=\color{slha_red}\ttfamily,
    keepspaces=true,
    showspaces=false,
    showstringspaces=false,
    rulecolor=\color{slha_frame},
    frame=true,
    frameround={t}{t}{t}{t},
    framexleftmargin=6mm,
    numbers=left,
    numberstyle=\tiny\color{halfgray},
    backgroundcolor=\color{slha_bg},
    basicstyle=\footnotesize,
    keywordstyle=\color{slha_green}\ttfamily,
    aboveskip=1.2em,
    belowskip=1.2em,
}

\lstdefinelanguage{SLHA}{
    morekeywords={block,Block,BLOCK,decay,Decay,DECAY},%
    sensitive=true,%
    morecomment=[l]\#,%
    morestring=[b]',%
    morestring=[b]",%
    morestring=[s]{'''}{'''},% used for documentation text (mulitiline strings)
    morestring=[s]{"""}{"""},% added by Philipp Matthias Hahn
    morestring=[s]{r'}{'},% `raw' strings
    morestring=[s]{r"}{"},%
    morestring=[s]{r'''}{'''},%
    morestring=[s]{r"""}{"""},%
    morestring=[s]{u'}{'},% unicode strings
    morestring=[s]{u"}{"},%
    morestring=[s]{u'''}{'''},%
    morestring=[s]{u"""}{"""},%
    identifierstyle=\color{black}\ttfamily,
    commentstyle=\color{slha_cyan}\ttfamily,
    stringstyle=\color{slha_red}\ttfamily,
    keepspaces=true,
    showspaces=false,
    showstringspaces=false,
    rulecolor=\color{slha_frame},
    frame=true,
    frameround={t}{t}{t}{t},
    framexleftmargin=6mm,
    numbers=left,
    numberstyle=\tiny\color{halfgray},
    backgroundcolor=\color{slha_bg},
    basicstyle=\footnotesize,
    keywordstyle=\color{slha_green}\ttfamily,
    aboveskip=1.2em,
    belowskip=1.2em,
}

\lstdefinestyle{terminal} {
  morekeywords={cp,-r,make,cd},
  numbers=left, 
  stepnumber=1, 
  numberstyle=\tiny\color{halfgray}, 
  numbersep=10pt, 
  backgroundcolor=\color{black}, 
  basicstyle=\color{white}\ttfamily,
  stringstyle=\color{white}\ttfamily,
  keywordstyle=\color{white}\ttfamily\bfseries
 }

\lstdefinelanguage{MathIn}{
    morekeywords={Simplify,Eigenvalues,epsUV,Delta,UVscaleQ},%
    emph={Start,InitUnitarity,GetScatteringDiagrams,BuildScatteringMatrix,MakeSPheno,InitMatching,EFTcoupLO,EFTcoupNLO},%
    emphstyle={\color{mathematica_purple}},
    sensitive=true,%
    morecomment=[l]\%,%
    morestring=[b]',%
    morestring=[b]",%
    morestring=[s]{'''}{'''},% used for documentation text (mulitiline strings)
    morestring=[s]{"""}{"""},% added by Philipp Matthias Hahn
    morestring=[s]{r'}{'},% `raw' strings
    morestring=[s]{r"}{"},%
    morestring=[s]{r'''}{'''},%
    morestring=[s]{r"""}{"""},%
    morestring=[s]{u'}{'},% unicode strings
    morestring=[s]{u"}{"},%
    morestring=[s]{u'''}{'''},%
    morestring=[s]{u"""}{"""},%
    identifierstyle=\color{black}\ttfamily,
    commentstyle=\color{mathematica_cyan}\ttfamily,
    stringstyle=\color{mathematica_red}\ttfamily,
    keepspaces=true,
    showspaces=false,
    showstringspaces=false,
    rulecolor=\color{mathematica_frame},
    frame=none,
    numbers=left,
    numberstyle=\tiny\color{halfgray},
    basicstyle=\footnotesize,
    keywordstyle=\color{mathematica_green}\ttfamily,
    aboveskip=0.2em,
    belowskip=0.2em
}

\lstdefinelanguage{MathOut}{
    morekeywords={Simplify,Eigenvalues},%
    sensitive=true,%
    morecomment=[l]\%,%
    morestring=[b]',%
    morestring=[b]",%
    morestring=[s]{'''}{'''},% used for documentation text (mulitiline strings)
    morestring=[s]{"""}{"""},% added by Philipp Matthias Hahn
    morestring=[s]{r'}{'},% `raw' strings
    morestring=[s]{r"}{"},%
    morestring=[s]{r'''}{'''},%
    morestring=[s]{r"""}{"""},%
    morestring=[s]{u'}{'},% unicode strings
    morestring=[s]{u"}{"},%
    morestring=[s]{u'''}{'''},%
    morestring=[s]{u"""}{"""},%
    identifierstyle=\color{black}\ttfamily,
    commentstyle=\color{mathematica_cyan}\ttfamily,
    stringstyle=\color{mathematica_red}\ttfamily,
    keepspaces=true,
    showspaces=false,
    showstringspaces=false,
    rulecolor=\color{mathematica_frame},
    frame=none,
    frameround={t}{t}{t}{t},
    framexleftmargin=10mm,
    numbers=left,
    numberstyle=\tiny\color{halfgray},
    basicstyle=\footnotesize,
    keywordstyle=\color{mathematica_green}\ttfamily,
    aboveskip=0.2em,
    belowskip=0.2em,
}

\let\origthelstnumber\thelstnumber
\makeatletter
\newcommand*\Suppressnumber{%
  \lst@AddToHook{OnNewLine}{%
    \let\thelstnumber\relax%
     \advance\c@lstnumber-\@ne\relax%
    }%
}

\newcommand*\Reactivatenumber{%
  \lst@AddToHook{OnNewLine}{%
   \let\thelstnumber\origthelstnumber%
   \advance\c@lstnumber\@ne\relax}%
}

\definecolor{maroon}{cmyk}{0, 0.87, 0.68, 0.32}
\definecolor{halfgray}{gray}{0.55}
\definecolor{ipython_frame}{RGB}{207, 207, 207}
\definecolor{ipython_bg}{RGB}{247, 247, 247}
\definecolor{ipython_red}{RGB}{186, 33, 33}
\definecolor{ipython_green}{RGB}{0, 128, 0}
\definecolor{ipython_cyan}{RGB}{64, 128, 128}
\definecolor{ipython_purple}{RGB}{170, 34, 255}

\lstdefinelanguage{iPython}{
    morekeywords={access,and,break,class,continue,def,del,elif,else,except,exec,finally,for,from,global,if,import,in,is,lambda,not,or,pass,print,raise,return,try,while},%
    morekeywords=[2]{abs,all,any,basestring,bin,bool,bytearray,callable,chr,classmethod,cmp,compile,complex,delattr,dict,dir,divmod,enumerate,eval,execfile,file,filter,float,format,frozenset,getattr,globals,hasattr,hash,help,hex,id,input,int,isinstance,issubclass,iter,len,list,locals,long,map,max,memoryview,min,next,object,oct,open,ord,pow,property,range,raw_input,reduce,reload,repr,reversed,round,set,setattr,slice,sorted,staticmethod,str,sum,super,tuple,type,unichr,unicode,vars,xrange,zip,apply,buffer,coerce,intern},%
    sensitive=true,%
    morecomment=[l]\#,%
    morestring=[b]',%
    morestring=[b]",%
    morestring=[s]{'''}{'''},% used for documentation text (mulitiline strings)
    morestring=[s]{"""}{"""},% added by Philipp Matthias Hahn
    morestring=[s]{r'}{'},% `raw' strings
    morestring=[s]{r"}{"},%
    morestring=[s]{r'''}{'''},%
    morestring=[s]{r"""}{"""},%
    morestring=[s]{u'}{'},% unicode strings
    morestring=[s]{u"}{"},%
    morestring=[s]{u'''}{'''},%
    morestring=[s]{u"""}{"""},%
    literate=
    {á}{{\'a}}1 {é}{{\'e}}1 {í}{{\'i}}1 {ó}{{\'o}}1 {ú}{{\'u}}1
    {Á}{{\'A}}1 {É}{{\'E}}1 {Í}{{\'I}}1 {Ó}{{\'O}}1 {Ú}{{\'U}}1
    {à}{{\`a}}1 {è}{{\`e}}1 {ì}{{\`i}}1 {ò}{{\`o}}1 {ù}{{\`u}}1
    {À}{{\`A}}1 {È}{{\'E}}1 {Ì}{{\`I}}1 {Ò}{{\`O}}1 {Ù}{{\`U}}1
    {ä}{{\"a}}1 {ë}{{\"e}}1 {ï}{{\"i}}1 {ö}{{\"o}}1 {ü}{{\"u}}1
    {Ä}{{\"A}}1 {Ë}{{\"E}}1 {Ï}{{\"I}}1 {Ö}{{\"O}}1 {Ü}{{\"U}}1
    {â}{{\^a}}1 {ê}{{\^e}}1 {î}{{\^i}}1 {ô}{{\^o}}1 {û}{{\^u}}1
    {Â}{{\^A}}1 {Ê}{{\^E}}1 {Î}{{\^I}}1 {Ô}{{\^O}}1 {Û}{{\^U}}1
    {œ}{{\oe}}1 {Œ}{{\OE}}1 {æ}{{\ae}}1 {Æ}{{\AE}}1 {ß}{{\ss}}1
    {ç}{{\c c}}1 {Ç}{{\c C}}1 {ø}{{\o}}1 {å}{{\r a}}1 {Å}{{\r A}}1
    {€}{{\EUR}}1 {£}{{\pounds}}1
    {^}{{{\color{ipython_purple}\^{}}}}1
    {=}{{{\color{ipython_purple}=}}}1
    {+}{{{\color{ipython_purple}+}}}1
    {*}{{{\color{ipython_purple}$^\ast$}}}1
    {/}{{{\color{ipython_purple}/}}}1
    {+=}{{{+=}}}1
    {-=}{{{-=}}}1
    {*=}{{{$^\ast$=}}}1
    {/=}{{{/=}}}1,
    literate=
    *{-}{{{\color{ipython_purple}-}}}1
     {?}{{{\color{ipython_purple}?}}}1,
    identifierstyle=\color{black}\ttfamily,
    commentstyle=\color{ipython_cyan}\ttfamily,
    stringstyle=\color{ipython_red}\ttfamily,
    keepspaces=true,
    showspaces=false,
    showstringspaces=false,
    rulecolor=\color{ipython_frame},
    frame=single,
    frameround={t}{t}{t}{t},
    framexleftmargin=6mm,
    numbers=left,
    numberstyle=\tiny\color{halfgray},
    backgroundcolor=\color{ipython_bg},
    basicstyle=\footnotesize,
    keywordstyle=\color{ipython_green}\ttfamily,
    aboveskip=1.2em,
    belowskip=1.2em,
}

\newcommand{\xSLHA}{\texttt{xSLHA}\xspace}

\newcommand{\SARAH}{\texttt{SARAH}\xspace}
\newcommand{\SPheno}{\texttt{SPheno}\xspace}

\begin{document}

\begin{frontmatter}

\title{%
\xSLHA:
an Les Houches Accord reader for Python and Mathematica
}

\author[kit1,kit2]{Florian Staub}

\address[kit1]{Institute for Theoretical Physics (ITP), Karlsruhe Institute of Technology, Engesserstra{\ss}e 7, D-76128 Karlsruhe, Germany}
\address[kit2]{Institute for Nuclear Physics (IKP), Karlsruhe Institute of Technology, Hermann-von-Helmholtz-Platz 1, D-76344 Eggenstein-Leopoldshafen, Germany}

\begin{abstract}
The format defined by the SUSY Les Houches Accord (SLHA) is widely used in high energy physics to store and exchange information. It is no longer applied only to a few supersymmetric models, but the general structure is adapted to all kind of models. Therefore, it is helpful to have parsers at hand which can import files in the SLHA format into high-level languages as Python and Mathematica in order to further process the data. The focus of the \xSLHA package, which exists now for Python and Mathematica, was on a fast read-in of large data samples. Moreover, also some blocks used by different tools, as {\tt HiggsBounds} for instance, deviate from the standard conventions. These are also supported by \xSLHA.
\end{abstract}

\end{frontmatter}

\section{Installation}
In phenomenological High Energy Physics (HEP) the heavy use of numerical tools is unavoidable to obtain reliable predictions. Consequently, very sophisticated tools for collider studies, mass spectra calculations, or the prediction of dark matter properties have been developed and are widely used by the community. Since the different tools often need to be linked, a common format to represent information is necessary. The so called 'SUSY Les Houches Accord' (SLHA) \cite{Skands:2003cj,Allanach:2008qq} proposed such a format in the context of minimal supersymmetric theories already 15 years ago. In the following years, there have been proposals have to extent the format to non-minimal supersymmetric models \cite{Basso:2012ew}, how to include information about flavour physics \cite{Mahmoudi:2010iz} or how to encode cross sections \cite{xsection}. Nowadays, the SLHA format with its block structure is not only used in the context of supersymmetric models, but for all kind of BSM models. Therefore, code has been written to parse files in the SLHA format into different programming languages like C \cite{lhpc,slhaea}, Fortran \cite{Hahn:2004bc,Hahn:2006nq}, Mathematica \cite{Marquard:2013ita} or Python \cite{Buckley:2013jua,pylha}. \\
These parsers usually assume single SLHA files as input. However, re-running them very often in order to read many files into high-level languages as Mathematica or Python for further processing the data can became very time and memory consuming. Therefore, another SLHA parser appears now on the market: \xSLHA for which a Python and Mathematica version exist. \xSLHA makes use of fast shell tools as {\tt cat} and {\tt grep} to prepare the data before reading it in. Moreover, it is the first (public) model independent SLHA reader for Mathematica, and supports also some non-standard blocks as used by {\tt HiggsBounds} or {\tt HiggsSignals} for instance.\\
This manual explains the usage of \xSLHA. 
In \cref{sec:python} it is shown how to install and use the Python version of \xSLHA. In \cref{sec:math} the same is done for Mathematica version. A brief summary is given in \cref{sec:summary}.

\section{\xSLHA for Python}
\label{sec:python}
\subsection{Installation}
The repository for the python version of \xSLHA as available at
\begin{center}
https://github.com/fstaub/xSLHA
\end{center}
However, the most convenient way to install it, is to make use of {\tt pip}.
The package is downloaded and installed by running 
\begin{lstlisting}[style=terminal]
> pip install xslha
\end{lstlisting}
in the terminal. Afterwards, it can be loaded in python by

\begin{lstlisting}[language=ipython]
import xslha
\end{lstlisting}

\subsection{Reading a single spectrum file}
Reading a spectrum file \texttt{file} and storing the information in a
class object \texttt{spc} is done via the command
\begin{lstlisting}[language=ipython]
spc=xslha.read(file)
\end{lstlisting}
in python. 
One has afterwards access to the different information by using the
\texttt{Value} command of the \texttt{xSLHA} class. 
\begin{lstlisting}[language=ipython]
spc.Value('Keyword', [Numbers])
\end{lstlisting}
For instance, common entries in
a MSSM spectrum file are extracted as
\begin{lstlisting}[language=ipython]
print("tan(beta): ",spc.Value('MINPAR',[3]))
print("T_u(3,3): ",spc.Value('TU',[3,3]))
print("m_h [GeV]: ",spc.Value('MASS',[25]))
print("Gamma(h) [GeV]: ",spc.Value('WIDTH',25))
print("BR(h->W^+W^-): ",spc.Value('BR',[25,[-13,13]]))
print("Sigma(pp->N1 N1,Q=8TeV): ",spc.Value('XSECTION',[8000,(2212,2212),(1000021,1000021)]))
\end{lstlisting}
This produces the following output:
\begin{lstlisting}[language=ipython]
tan(beta):  16.870458
T_u(3,3):  954.867627
m_h [GeV]:  117.758677
Gamma(h) [GeV]:  0.00324670136
BR(h->W^+W^-):  0.000265688227
Sigma(pp->N1 N1,Q=8TeV): [[(0, 2, 0, 0, 0, 0), 0.00496483158]]
\end{lstlisting}
Thus, the conventions are: 
\begin{itemize}
\item the information given in the different SLHA
blocks is returned by using using the name of the block as well as the corresponding number in the block
as input
\item the widths of particles are
returned via the keyword \texttt{WIDTH} and giving the  PDG of the particle 
\item for branching ratios, the keyword \texttt{BR} is used together with a
nested list which states the PDGs of the decaying particle as well as of the
final states 
\item for cross-sections the keyword \texttt{XSECTION} is used
together with a nested list which states the centre-of-mass energy and
the PDGs of the initial/final states. The result is a list containing
all calculated cross-sections for the given options for the
renormalisation scheme, the QED \& QCD order, etc. (see the SLHA
recommendations for details).
\end{itemize}

Another possibility to access the information in the spectrum file is to
look at the different dictionaries which are created by the read command\footnote{We comment later on {\tt widths1L} and {\tt br1L}.}:
\begin{lstlisting}[language=ipython]
spc.blocks
spc.widths
spc.br
spc.widths1L
spc.br1L
spc.xsctions
\end{lstlisting}
These dictionaries contain all information stored in the SLHA file. 

\subsection{Reading all spectrum files from a directory}
In order to read several spectrum files located in a directory
\texttt{dir}, one can make use of the command
\begin{lstlisting}[language=ipython]
list_spc=xslha.read_dir(dir)
\end{lstlisting}
This generates a list \texttt{list\_spc} where each entry corresponds to
one spectrum file. Thus, one can for instance use
\begin{lstlisting}[language=ipython]
[[x.Value('MINPAR',[1]),x.Value('MASS',[25])] for x in list_spc]
\end{lstlisting}
to extract the input for a 2D-scatter plot which shows the dependence of $m_h$ (\texttt{MASS[25]}) on $m_0$ (\texttt{MINPAR[1]}).

\subsection{Fast read-in of many files}
\label{sec:python_read_dir}
Reading many spectrum files can be time consuming. However, many of the
information which is given in a SLHA file is often not needed for a
current study. Therefore, one can speed up the reading by extracting
first all relevant information. This generates smaller files which are
faster to read in. This can be done via the optional argument
\texttt{entries} for \texttt{read\_dir}:

\begin{lstlisting}[language=ipython]
list_spc_fast=xslha.read_dir("/home/$USER/Documents/spc1000",entries=["# m0","# m12","# hh_1"])`
\end{lstlisting}
\texttt{entries} defines a list of strings which can be used to extract
the necessary lines from the SLHA file by using \texttt{cat} and \texttt{grep}. Usually, the comments which are given for most entries in a SLHA file 
are very suitable for this purpose. For instance, the spectrum file for a CMSSM point generated with \SARAH version of \SPheno looks like 
\begin{lstlisting}[language=SLHA]
Block MINPAR  # Input parameters
    1    1.50000000E+03  # m0
    2    1.50000000E+03  # m12
    3    1.00000000E+01  # TanBeta
    4    1.00000000E+00  # SignumMu
    5   -2.00000000E+03  # Azero
...
Block MASS  # Mass spectrum
#   PDG code      mass          particle
...
        25     1.21048604E+02  # hh_1
        35     2.63360770E+03  # hh_2
        36     2.63354615E+03  # Ah_2
\end{lstlisting}
This explains, why we have chosen \verb'entries=["# m0","# m12","# hh_1"]' in the example.  \\

The speed improvement can be easily an order of magnitude if only some entries
from a SLHA file are actually needed. We demonstrate this at a short example reading 1,000 SLHA files. \\
Reading the full spectrum files takes about 5--6 seconds:
\begin{lstlisting}[language=ipython]
%%time
list_spc=xslha.read_dir("/home/$USER/Documents/spc1000")

CPU times: user 5.05 s, sys: 105 ms, total: 5.15 s
Wall time: 5.51 s
\end{lstlisting}
while the relevant information is read in less than one second:
\begin{lstlisting}[language=ipython]
%%time
list_spc_fast=xslha.read_dir("/home/$USER/Documents/spc1000",entries=["# m0","# m12","# hh_1"])

CPU times: user 147 ms, sys: 132 ms, total: 280 ms
Wall time: 917 ms
\end{lstlisting}

We can compare these numbers also with the running time when using other available python parsers to read in all files
\begin{itemize}
\item \texttt{pylha}:
\begin{lstlisting}[language=ipython]
%%time
all_spc=[]
for filename in os.listdir("/home/$USER/Documents/spc1000/"): 
  with open("/home/$USER/Documents/spc1000/"+filename) as f:
    input=f.read()
    all_spc.append(pylha.load(input))
    
CPU times: user 21.5 s, sys: 174 ms, total: 21.7 s
Wall time: 21.7 s    
\end{lstlisting}

\item \texttt{pyslha}
\begin{lstlisting}[language=ipython]
%%time
all_spc=[]
for filename in os.listdir("/home/$USER/Documents/spc1000/"): 
    all_spc.append(pyslha.read("/home/$USER/Documents/spc1000/"+filename))
    
CPU times: user 13.3 s, sys: 152 ms, total: 13.5 s Wall time: 13.5 s
\end{lstlisting}
\end{itemize}
We see that \xSLHA even without any restriction is already significantly faster for a large sample of files compared to the other packages.

\subsection{Reading several spectra stored in a single file}
Another common approach for saving spectrum files is to produce one huge
file in which the different spectra are separated by a keyword.
\texttt{xSLHA} can read such files by setting the optional argument
\texttt{separator} for \texttt{read}:
\begin{lstlisting}[language=ipython]
list_spc=xslha.read(file,separator=keyword)
\end{lstlisting}

In order to speed up the reading of many spectra also in this case, it
is possible to produce first a smaller version of the files by defining the entries 
which should be read. The command for this reads
\begin{lstlisting}[language=ipython]
list_spc=xslha.read_small(file,entries,separator)
\end{lstlisting}
In this case \texttt{xSLHA} extract all relevant lines first
using \texttt{cat} and \texttt{grep} again. For instance, in order to read
efficiently large files produced with \texttt{SSP} \cite{Staub:2011dp}, one can use:
\begin{lstlisting}[language=ipython]
list_spc=xslha.read("SpectrumFiles.spc",["# m0", "# m12", "# hh_1"],"ENDOFPARAMETERFILE")
\end{lstlisting}

\subsection{Special blocks}
\label{sec:special-blocks}
\label{sec:special}
There are some programs which use blocks that are not supported by the
official SLHA conventions: 
\begin{itemize}
\item \texttt{HiggsBounds}\cite{Bechtle:2008jh,Bechtle:2011sb,Bechtle:2013wla} expects
the effective coupling ratios in the blocks
\begin{itemize}
\item \texttt{HIGGSBOUNDSINPUTHIGGSCOUPLINGSBOSONS} 
\item \texttt{HIGGSBOUNDSINPUTHIGGSCOUPLINGSFERMIONS} 
\end{itemize}
which are differently
ordered compared to other blocks:
\begin{lstlisting}[language=SLHA]
Block HiggsBoundsInputHiggsCouplingsFermions # 
 6.67667670E-01    0.00000000E+00  3   25 5 5 # h_1 b b coupling 
 5.66931862E-01    0.00000000E+00  3   25 3 3 # h_1 s s coupling 
...
Block HiggsBoundsInputHiggsCouplingsBosons # 
 1.11812190E+00    3   25 24 24 # h_1 W W coupling 
 1.08018976E+00    3   25 23 23 # h_1 Z Z coupling  
\end{lstlisting}
Thus, first the numerical entries are stated
before the number and the PDGs of the involved particles follow
\item \texttt{SPheno}\cite{Porod:2003um,Porod:2011nf} version generated by
\texttt{SARAH}\cite{Staub:2008uz,Staub:2009bi,Staub:2010jh,Staub:2012pb,Staub:2013tta} can calculate one-loop
corrections to the decays. The results are given in the blocks
\texttt{DECAY1L} which appear in parallel to \texttt{DECAY} containing
the standard calculation: 
\begin{lstlisting}[language=SLHA]
DECAY   1000001     5.02288570E+01   # Sd_1
#    BR                NDA      ID1      ID2
     2.84234837E-01    2        6   -1000024   # BR(Sd_1 -> Fu_3 Cha_1 )
     1.85835591E-01    2        6   -1000037   # BR(Sd_1 -> Fu_3 Cha_2 )
     2.20528398E-04    2        3    1000023   # BR(Sd_1 -> Fd_2 Chi_2 )
....

DECAY1L   1000001     4.79764029E+01   # Sd_1
#    BR                NDA      ID1      ID2
     2.66279140E-01    2        6   -1000024   # BR(Sd_1 -> Fu_3 Cha_1 )
     1.59538386E-01    2        6   -1000037   # BR(Sd_1 -> Fu_3 Cha_2 )
     2.25120157E-04    2        3    1000023   # BR(Sd_1 -> Fd_2 Chi_2 )
\end{lstlisting}

\texttt{xSLHA} will distinguish these cases
when reading the file and offer the two following options for
\texttt{Values} in addition:
\begin{lstlisting}[language=ipython]
spc.Values('WIDTH1L',1000022)
spc.Values('BR1L',[1000023,[25,1000022]])
\end{lstlisting}
\end{itemize}

% \subsection{Writing files}\label{writing-files}
% Files in the SLHA format can be written via
% 
% \begin{lstlisting}[language=ipython]
% xslha.write(blocks,file)
% \end{lstlisting}
% 
% where it might be the best to use ordered dictionaries to define the
% blocks and the values in the blocks. For instance
% 
% \begin{lstlisting}[language=ipython]
% import collections
% out_blocks=collections.OrderedDict([
%               ('MODSEL',collections.OrderedDict([('1', 1), ('2', 2),('6',0)])),
%               ('MINPAR',collections.OrderedDict([('1', 1000.),('2', 2000),('3',10),('4',1),('5',0)]))
% ])
% xslha.write(out_blocks,"/home/$USER/Documents/LH.in")
% \end{lstlisting}

\section{\xSLHA for Mathematica}
\label{sec:math}
The Mathematica version of \xSLHA is a re-write of the SLHA parser included in {\tt SSP}. 
It is now very similar in its functionality to the Python version of \xSLHA. Thus, the reader 
who carefully went through the last section will recognise quite some repetition in this section. This was kept 
in order to have also a stand-alone manual for the {\tt Mathematica} version for those who don't want to go through
the Python section. 

\subsection{Installation}\label{installation}
The repository for the Mathematica version of \xSLHA is available at github:
\begin{center}
{\tt github.com/fstaub/xSLHA.m} 
\end{center}
The simplest way to install the package is to put the file \texttt{xSLHA.m}
into a {\tt xSLHA} sub-directory of the applications directory of \texttt{Mathematica}. Thus,
\begin{center}
/home/\$USER/.Mathematica/Applications/xSLHA/  
\end{center}
is the best place to store {\tt xSLHA.m}. Afterwards, the
package can be loaded via
\begin{MIN}
<<xSLHA`
\end{MIN}
in Mathematica. 

\subsection{Reading a single spectrum file}
Reading a spectrum file \texttt{file} and storing the information in a
variable \texttt{spc} is done via the command
\begin{MIN}
spc=xSLHA`Read[file]
\end{MIN}
The content of the spectrum file is returned on form of a list of
replacements. Those can be used as follows in order to extract specific information:
\begin{MIN}
Print["tan(beta): ",MINPAR[3]/.spc]
Print["T_u(3,3): ",TU[3,3]/.spc]
Print["m_h [GeV]: ",MASS[25]/.spc]
Print["Gamma(h) [GeV]: ",WIDTH[25]/.spc]
Print["BR(h->W^+W^-): ",BR[25][-13,13]/.spc]
Print["Sigma(pp->N1 N1,Q=8TeV): ",XSECTION[8000,[2212,2212],[1000021,1000021]]/.spc]
\end{MIN}
This produces the following output
\begin{MOUT}
tan(beta):  16.870458
T_u(3,3):  954.867627
m_h [GeV]:  117.758677
Gamma(h) [GeV]:  0.00324670136
BR(h->W^+W^-):  0.000265688227
Sigma(pp->N1 N1,Q=8TeV): {{{0, 2, 0, 0, 0, 0}, 0.00496483158}}
\end{MOUT}
Thus, the conventions are: 
\begin{itemize}
\item the information given in the different SLHA
blocks is returned by using using the name of the block as well as the corresponding number in the block
as input
\item the widths of particles are
returned via the keyword \texttt{WIDTH} and giving the  PDG of the particle 
\item for branching ratios, the keyword \texttt{BR} is used together with two
sets of arguments which state the PDG of the decaying particle as well
as of the final states
\item for cross-sections the keyword \texttt{XSECTION} is used
together with a nested list which states the centre-of-mass energy and
the PDGs of the initial/final states. The result is a list containing
all calculated cross-sections for the given options for the
renormalisation scheme, the QED \& QCD order, etc. (see the SLHA
recommendations for details).
\end{itemize}

\subsection{Reading all spectrum files from a directory}
In order to read several spectrum files located in a directory
\texttt{dir}, one can make use of the command
\begin{MIN}
listSPC=xSLHA`ReadDir[dir]
\end{MIN}
This generates a list \texttt{listSPC} where each entry corresponds to
one spectrum. Thus, one can for instance use
\begin{MIN}
{MINPAR[1],MASS[25]}/. listSPC
\end{MIN}
to extract the input for a 2D-scatter plot showing the dependence of $m_h$ (\texttt{MASS[25]}) on $m_0$ (\texttt{MINPAR[1]}) in a CMSSM scan.

\subsection{Fast read-in of many files}
Reading many spectrum files can be very time consuming -- especially in Mathematica. However, many of the
information which is given in a SLHA file is often not needed for the
current study. Therefore, one can speed up the reading by extracting
first all relevant information. This generates smaller files which are
faster to read. This can be done via the optional argument
\texttt{entries} for \texttt{ReadDir}:
\begin{MIN}
ListSpcFast=xSLHA`ReadDir["/home/$USER/Documents/spc1000/",entries={"# m0","# m12","# hh_1"}]`
\end{MIN}
\texttt{entries} defines a list of strings which can be used to extract
the necessary lines from the SLHA file by using \texttt{cat} and \texttt{grep}. Usually, the comments which are given for most entries in a SLHA file 
are very suitable for this purpose. Snippets of a CMSSM point generated with \SPheno are shown in \cref{sec:python_read_dir}.
This explains, why we have chosen \verb'entries=["# m0","# m12","# hh_1"]' in the example.  \\

% \subsubsection{Speed}\label{speed}

The impact of this optimisation for reading 1,000 files is as follows. Reading the full files takes about 3 minutes
\begin{MIN}
Timing[
 ListSPC = xSLHA`ReadDir["/home/$USER/Documents/spc1000/"];
 ]

{158.996, Null}
\end{MIN}
This needs to be compared with the 6--7 seconds needed to read the smaller files:
\begin{MIN}
Timing[
 ListSPC = 
   xSLHA`ReadDir["/home/$USER/Documents/spc1000/", 
    entries -> {"# m0", "# m12", "# hh_1"}];
 ]

{6.48319, Null}
\end{MIN}
We see that Mathematica is significantly slower in reading these files compared to Python. Thus, pre-processing the files is absolutely necessary for dealing with large data samples in Mathematica. 

\subsection{Reading several spectra stored in a single file}
Another common approach for saving spectrum files is to produce one huge
file in which the different spectra are separated by a keyword.
\texttt{xSLHA} can read such files by setting the optional argument
\texttt{separator} for \texttt{Read}:

\begin{MIN}
listSPC=xSLHA`Read[file,separator->string]
\end{MIN}
In order to speed up the reading of many spectra also in this case, it
is possible to define the entries as well which are need:
\begin{MIN}
listSPC=xSLHA`ReadSmall[file,separator->string,entries->list]
\end{MIN}
In this case\texttt{xSLHA} will produce first a smaller spectrum file
using \texttt{cat} and \texttt{grep}. For instance, in order to read
efficiently files produced with
\texttt{SSP}, one can use:
\begin{MIN}
listSPC=xSLHA`ReadSmall["SpectrumFiles.spc",separator->"ENDOFPARAMETERFILE",entries->{"# m0", "# m12", "# hh_1"}]
\end{MIN}

\subsection{Special blocks}
There are some programs which use blocks that are not supported by the
official SLHA conventions. We have collected them already in \cref{sec:special-blocks} where also a bit more information
about them is given.
Those are also supported by \xSLHA and are accessible via:
\begin{itemize}
 \item {\tt HiggsBounds} blocks:
\begin{MIN}
HIGGSBOUNDSINPUTHIGGSCOUPLINGSBOSONS[25,22,22] /.spc
HIGGSBOUNDSINPUTHIGGSCOUPLINGSFERMIONS[25,13,-13] /.spc
\end{MIN}
 \item One-loop decays:
\begin{MIN}
WIDTH1L[1000022] /.spc
BR1L[1000023][25,1000022] /.spc
\end{MIN}
\end{itemize}

\section{Summary}
\label{sec:summary}
This manual introduces the SLHA parser \xSLHA for Python and Mathematica. \xSLHA is a handy and flexible reader which makes it easy to use the data stored in a SLHA spectrum file. It also provides wrapper functions to extract the relevant information from the files by using shell tools before reading it in. This can speeds up the important of large data samples by an order of magnitude and more.

\section*{Acknowledgements}
I thank Martin Gabelmann Toby Opferkuch for testing \xSLHA under Linux and MacOS. 
This work is supported by the ERC Recognition Award ERC-RA-0008 of the Helmholtz Association.

\bibliographystyle{ArXiv}
\bibliography{lit.bib}

\end{document}